\documentclass[letterpaper,english,aps,letter,prl,twocolumn,superscriptaddress]{revtex4}
\usepackage[T1]{fontenc}
\usepackage[latin1]{inputenc}
\usepackage{amsmath}
\usepackage{graphicx}
\usepackage{amssymb}

\makeatletter

\usepackage{ae,aecompl}
\renewcommand{\citet}[1]{\cite{#1}}

\usepackage{babel}
\makeatother
\begin{document}

\title{Stochastic Hard-Sphere Dynamics for Hydrodynamics of Non-Ideal Fluids}

\author{Aleksandar Donev}

\affiliation{Lawrence Livermore National Laboratory, P.O.Box 808, Livermore, CA
94551-9900}

\author{Berni J. Alder}

\affiliation{Lawrence Livermore National Laboratory, P.O.Box 808, Livermore, CA
94551-9900}

\author{Alejandro L. Garcia}

\affiliation{Department of Physics, San Jose State University, San Jose, California,
95192}

\begin{abstract}
A novel stochastic fluid model is proposed with non-ideal structure
factor consistent with compressibility, and adjustable transport coefficients.
This Stochastic Hard Sphere Dynamics (SHSD) algorithm is a modification
of the Direct Simulation Monte Carlo (DSMC) algorithm and has several
computational advantages over event-driven hard-sphere molecular dynamics.
Surprisingly, SHSD results in an equation of state and pair correlation
function identical to that of a deterministic Hamiltonian system of
penetrable spheres interacting with linear core pair potentials. The
fluctuating hydrodynamic behavior of the SHSD fluid is verified for
the Brownian motion of a nano-particle suspended in a compressible
solvent.
\end{abstract}
\maketitle
\newcommand{\Cross}[1]{\left|\mathbf{#1}\right|_{\times}}
\newcommand{\CrossL}[1]{\left|\mathbf{#1}\right|_{\times}^{L}}
\newcommand{\CrossR}[1]{\left|\mathbf{#1}\right|_{\times}^{R}}
\newcommand{\CrossS}[1]{\left|\mathbf{#1}\right|_{\boxtimes}}

\newcommand{\V}[1]{\mathbf{#1}}
\newcommand{\M}[1]{\mathbf{#1}}
\newcommand{\D}[1]{\Delta#1}

\newcommand{\sV}[1]{\boldsymbol{#1}}
\newcommand{\sM}[1]{\boldsymbol{#1}}

\newcommand{\grad}{\boldsymbol{\nabla}}
\newcommand{\eij}{\left\{  i,j\right\}  }

\newcommand{\Wi}{\mbox{Wi}}

\newcommand{\modified}[1]{#1}

\newcommand{\deleted}[1]{}

\newcommand{\added}[1]{#1}

With the increased interest in nano- and micro-fluidics, it has become
necessary to develop tools for hydrodynamic calculations at the atomistic
scale \citet{ParticleMesoscaleHydrodynamics,FluctuatingHydro_Coveney}.
Of particular interest is the modeling of flexible polymers in a flowing
solvent for both biological (e.g., cell membranes) and engineering
(e.g., micro-channel DNA arrays) applications. Typically the polymer
chains are modeled using Molecular Dynamics (MD). For many applications,
a realistic representation of the solvent and bidirectional coupling
between the flow and the polymer motion is needed, for example, in
the modeling of turbulent drag reduction. Previously, we introduced
the Stochastic Event-Driven MD (SEDMD) algorithm that uses Direct
Simulation Monte Carlo (DSMC) for the solvent coupled to deterministic
EDMD for the polymer chain \citet{DSMC_AED}. However, DSMC is limited
to perfect gases. Efforts have been undertaken to develop solvents
that have a \emph{non-ideal} EOS, and that also have greater computational
efficiency than brute-force molecular dynamics. Examples include the
Lattice-Boltzmann (LB) method \citet{NonIdeal_LB}, Dissipative Particle
Dynamics (DPD) \citet{DPD_NonIdeal}, and Multi-Particle Collision
Dynamics (MPCD) \citet{MPCD_CBA}, each of which has its own advantages
and disadvantages \citet{ParticleMesoscaleHydrodynamics}. The \emph{Stochastic
Hard Sphere Dynamics} (SHSD) algorithm described in this Letter is
based on successive stochastic collisions of variable hard-sphere
diameters and is thermodynamically consistent (i.e., the direct calculation
of compressibility from density fluctuations agrees with the density
derivative of pressure). SHSD modifies previous algorithms for solving
the Enskog kinetic equation \citet{DSMC_Enskog_Frezzotti,DSMC_Enskog}
while maintaining good efficiency. 

In the SHSD algorithm randomly chosen pairs of approaching particles
that lie less than a given diameter of each other undergo collisions
as if they were hard spheres of diameter equal to their actual separation.
The SHSD fluid is shown to be non-ideal, with structure and equation
of state equivalent to that of a fluid mixture where spheres effectively
interact with a repulsive linear core pairwise potential. We theoretically
demonstrate this correspondence at low densities. Remarkably, we numerically
find that this effective interaction potential, similar to the quadratic
core potential used in many DPD variants, is valid at all densities.
\modified{Therefore, the SHSD fluid, as DPD, is \emph{intrinsically}
thermodynamically-consistent, while non-ideal MPCD is only \emph{numerically}
thermodynamically-consistent for tuned choices of the parameters \citet{MPCD_CBA,MPCD_CBA_2}.}

\modified{As an algorithm, SHSD is similar in nature to DPD and has
a similar computational complexity. In DPD, momentum is also stochastically
exchanged between particles closer than a given distance. The essential
difference is that DPD has a continuous-time formulation (a system
of stochastic ODEs), where as the SHSD dynamics is discontinuous in
time. This is similar to the difference between MD for continuous
potentials and discontinuous potentials. Just as DSMC is a stochastic
alternative to hard-sphere MD for low-density gases, SHSD is a stochastic
modification of hard-sphere MD for dense gases. On the other hand,
DPD is a modification of MD for smooth potentials to allow for larger
time-steps and a hydrodynamically-consistent thermostat.}

The SHSD algorithm is not as efficient as DSMC at a comparable collision
rate. However, when low compressibility is desired, SHSD is several
times faster than EDMD for hard spheres, the fastest available deterministic
alternative. Low compressibility, for example, is desirable so that
flows are kept subsonic even for high Reynolds number flows. Furthermore,
SHSD has several important advantages over EDMD, in addition to its
simplicity: (1) SHSD has several controllable parameters that can
be used to change the transport coefficients and compressibility,
while EDMD only has density; (2) SHSD is time-driven rather than event-driven
thus allowing for easy parallelization; (3) SHSD can be more easily
coupled to continuum hydrodynamic solvers, just like ideal-gas DSMC
\citet{FluctuatingHydro_AMAR}. \added{Strongly-structured particle
systems, such as fluids with strong interparticle repulsion (e.g.,
hard spheres), are more difficult to couple to hydrodynamic solvers
\citet{FluctuatingHydroHybrid_MD} than ideal fluids, such as MPCD
or DSMC, or weakly-structured fluids, such as DPD or SHSD fluids.}

The standard DSMC \citet{DSMCReview_Garcia} algorithm starts with
a time step where particles are propagated advectively, $\V{r}_{i}^{'}=\V{r}_{i}+\V{v}_{i}\D{t}$,
and sorted into a grid of cells. Then, a certain number $N_{coll}\sim\Gamma_{sc}N_{c}(N_{c}-1)\D{t}$
of \emph{stochastic collisions} are executed between pairs of particles
randomly chosen from the $N_{c}$ particles inside the cell. The conservative
stochastic collisions exchange momentum and energy between two particles
$i$ and $j$ that is not correlated with the actual positions of
the particles. Typically the probability of collision is made proportional
to the magnitude of the relative velocity $v_{r}=\left|\V{v}_{ij}\right|$
by using a conventional rejection procedure. \deleted{For mean free
paths comparable to the cell size, the grid of cells should be shifted
randomly before each collision step to ensure Galilean invariance.
}DSMC, unlike MD, is not microscopically isotropic and does not conserve
angular momentum, leading to an anisotropic stress tensor. To avoid
such grid artifacts, all collision partners within a collision diameter
$D$ must be considered even if they are in neighboring cells, and,
if angular momentum conservation is required, only radial momentum
should be exchanged in collisions as for hard spheres. This grid-free
variant will be called Isotropic DSMC (I-DSMC). The cost is that is
the computational efficiency is reduced by a factor of $2-3$ due
to the need to perform neighbor searches. \added{Note that a pairwise
Anderson thermostat proposed within the context of MD/DPD in Ref.
\citet{DPDSchmidtNumbers} essentially adds (thermostated) I-DSMC
collisions to ordinary MD and has very similar computational behavior.}
\modified{As in I-DSMC, in SHSD we consider particles in neighboring
cells as collision partners in order to ensure isotropy of the collisional
(non-ideal) component of the pressure tensor.}

The virial $\left\langle \D{\V{v}_{ij}}\cdot\D{\V{r}_{ij}}\right\rangle $
vanishes in I-DSMC giving an ideal-gas pressure. In order to introduce
a non-trivial equation of state it is necessary to either give an
additional displacement to the particles that is parallel to $\D{\V{v}_{ij}}$,
or to bias the momentum exchange $\D{\V{v}_{ij}}$ to be (statistically)
aligned to $\D{\V{r}_{ij}}$. The former approach has already been
investigated in the Consistent Boltzmann Algorithm (CBA) \citet{DSMC_CBA};
however, CBA is not thermodynamically consistent since it modifies
the compressibility without affecting the density fluctuations (i.e.,
the structure of the fluid is still that of a perfect gas). A fully
consistent approach is to require that the particles collide as if
they are elastic hard spheres of diameter equal to the distance between
them at the time of the collision. Such collisions produce a positive
virial only if the particles are approaching each other, $v_{n}=-\V{v}_{ij}\cdot\hat{\V{r}}_{ij}>0$,
therefore, we reject collisions among particles that are moving apart.
Furthermore, as for hard spheres, it is necessary to collide pairs
with probability that is \emph{linear} in $v_{n}$, which requires
a further increase of the rejection rate and thus decrease of the
efficiency. Without rejection based on $v_{n}$ or $v_{r}$, fluctuations
of the local temperature $T_{c}$ would not be consistently coupled
to the local pressure $p_{c}\sim\left\langle \D{\V{v}_{ij}}\cdot\D{\V{r}_{ij}}\right\rangle _{c}\sim\Gamma_{sc}\sqrt{T_{c}}$
because $p_{c}$ would be $\sim\sqrt{T_{c}}$ instead of the necessary
$p_{c}\sim T_{c}$. \deleted{This is because the EOS of a fluid with
no internal energy is linear in temperature, which from the virial
theorem $p_{c}\sim\left\langle \D{\V{v}_{ij}}\cdot\D{\V{r}_{ij}}\right\rangle _{c}\sim\Gamma_{sc}\sqrt{T_{c}}$
implies that the \emph{local} collisional frequency $\Gamma_{sc}$
must be proportional to the square root of the \emph{local} temperature
$T_{c}$, $\Gamma_{sc}\sim\sqrt{T_{c}}$, as for hard spheres.}For
DSMC the collisional rules can be manipulated arbitrarily to obtain
the desired transport coefficients, however, for non-ideal fluids
thermodynamic requirements eliminate some of the freedom. This important
observation has not been taken into account in other algorithms that
randomize hard-sphere MD \citet{PPM_DSMC}. Note that one can in fact
add I-DSMC collisions to SHSD in order to tune the viscosity without
affecting the compressibility. \deleted{The efficiency is significantly
enhanced when the fraction of accepted collisions is increased, however,
the compressibility is also increased at a comparable collision rate.}

For sufficiently small time steps, the SHSD fluid can be considered
as a simple modification of the standard hard-sphere fluid. Particles
move ballistically in-between collisions. When two particles $i$
and $j$ are less than a diameter apart, $r_{ij}\leq D$, there is
a probability rate $(3\chi/D)v_{n}\Theta(v_{n})$ for them to collide
as if they were elastic hard spheres with a variable diameter $D_{S}=r_{ij}$.
Here $\Theta$ is the Heaviside function, and $\chi$ is a dimensionless
parameter determining the collision frequency. The prefactor $3/D$
has been chosen so that for an ideal gas the average collisional rate
would be $\chi$ times larger than that of a low-density hard-sphere
gas with density (volume fraction) $\phi=\pi ND^{3}/(6V)$.

\begin{figure}[tbph]
\includegraphics[width=0.85\columnwidth,keepaspectratio]{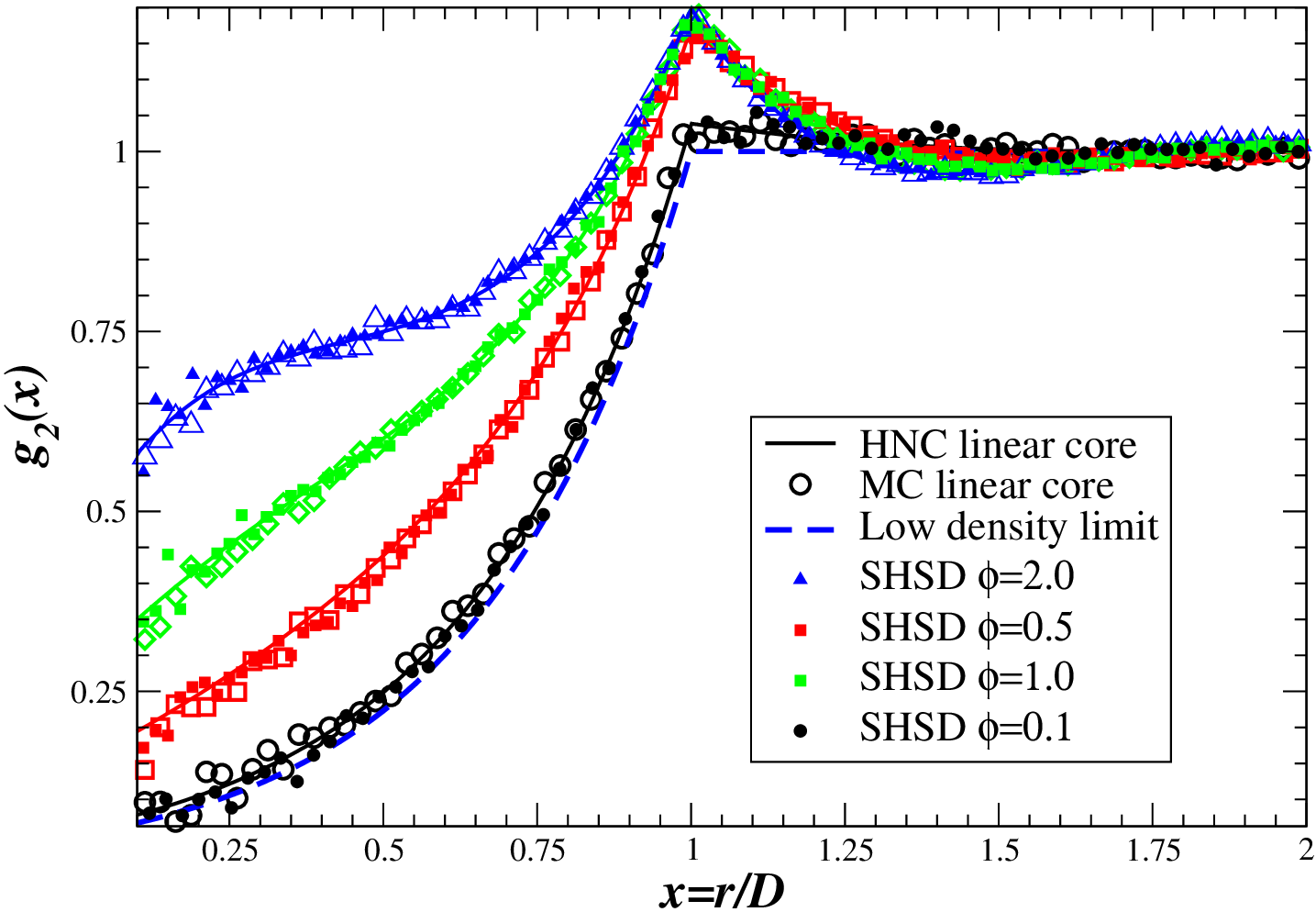}

\caption{\label{SHSD_g2}(Color online) Equilibrium pair correlation function
of the SHSD fluid (solid symbols), compared to MC (open symbols) and
HNC calculations (solid lines) for the linear core system, at various
densities and $\chi=1$.}
\end{figure}

In order to understand properties of the SHSD fluid as a function
of $\phi$ and $\chi$, we consider the equilibrium pair correlation
function $g_{2}$ at low densities, where correlations higher than
pairwise can be ignored. We consider the cloud of point walkers $ij$
representing the $N(N-1)/2$ pairs of particles, each at position
$\V{r}=\V{r}_{i}-\V{r}_{j}$ and with velocity $\V{v}=\V{v}_{i}-\V{v}_{j}$.
At equilibrium, the distribution of the point walkers in phase space
will be \emph{$f(\V{v},\V{r})=f(v_{r},r)\sim g_{2}(r)\exp(-mv_{n}^{2}/4kT)$.}
Inside the core $r<D$ this distribution of pair walkers satisfies
a kinetic equation\[
\frac{\partial f}{\partial t}+v_{n}\frac{\partial f}{\partial r}=v_{n}\Gamma_{0}f,\]
where $\Gamma_{0}=3\chi/D$ is the collision frequency. At equilibrium,
$\partial f/\partial t=0$ and $v_{n}$ cancels, consistent with choosing
collision probability linear in $\left|v_{n}\right|$. Thus $dg_{2}/dx=3\chi g_{2}\Theta(1-x),$
with solution $g_{2}(x)=\exp\left[3\chi(x-1)\right]$ for $x\leq1$
and $g_{2}(x)=1$ for $x>1$, where $x=r/D$. Indeed, numerical experiments
confirmed that at sufficiently low densities the equilibrium $g_{2}$
for the SHSD fluid has this exponential form inside the collision
core. This low density result is equivalent to $g_{2}^{U}=\exp[-U(r)/kT]$,
where $U(r)/kT=3\chi(1-x)\Theta(1-x)$ is an effective \emph{linear
core} pair potential similar to the quadratic core potential used
in DPD. Remarkably, it was found \emph{numerically} that this repulsive
potential can predict exactly $g_{2}(x)$ at \emph{all} liquid densities.
Figure \ref{SHSD_g2} shows a comparison between the pair correlation
function of the SHSD fluid on one hand, and a Monte Carlo calculation
using the linear core pair potential on the other, at several densities.
Also shown is a numerical solution to the hyper-netted chain (HNC)
integral equations for the linear core system\added{, inspired by
its success for the Gaussian core model \citet{GaussianCoreHNC}}.
The excellent agreement at all densities permits the use of the HNC
result in practical applications, notably the calculation of the transport
coefficients.

Interestingly, in the limit $\chi\rightarrow\infty$ the SHSD algorithm
reduces to hard-sphere (HS) molecular dynamics. In fact, if the density
$\phi$ is smaller than the freezing point for the HS system, the
structure of the SHSD fluid approaches, as $\chi$ increases, that
of the HS fluid. For higher densities, if $\chi$ is sufficiently
high, crystallization is observed in SHSD, either to the usual hard-sphere
crystals if $\phi$ is lower than the close-packing density, or if
not, to an unusual partially ordered state with multiple occupancy
per site, \added{typical of weakly repulsive potentials.}

An exact BBGKY-like hierarchy of Master equations for the $s$-particle
distribution functions of the SHSD fluid is given in Ref. \citet{StochasticFluid_Euler}.
For the first equation of this BBGKY hierarchy, valid at low densities,
we can neglect correlations other than pair ones and approximate $f_{2}(\V{r}_{1},\V{v}_{1},\V{r}_{2},\V{v}_{2})=g_{2}(\V{r}_{12})f_{1}(\V{r}_{1},\V{v}_{1})f(\V{r}_{2},\V{v}_{2})$.
With this assumption we obtain a stochastic Enskog equation similar
to a revised Enskog equation for hard spheres but with a smeared distribution
of hard-sphere diameters, as studied in Ref. \citet{VariableEnskog_Transport}.
The Chapman-Enskog expansion carried out in Ref. \citet{VariableEnskog_Transport}
produces the equation of state (EOS) $p=PV/NkT$, and approximations
to the self-diffusion coefficient $\zeta$, the shear $\eta$ and
bulk $\eta_{B}$ viscosities, and thermal conductivity $\kappa$ of
the SHSD fluid. The expressions ultimately give the transport coefficients
in terms of various integer moments of $g_{2}(x)$, $x_{k}=\int_{0}^{1}x^{k}g_{2}(x)dx$,
specifically, $p-1=12\phi\chi x_{3}$, $\zeta/\zeta_{0}=\sqrt{\pi}/(48\phi\chi x_{2})$,
$\eta_{B}/\eta_{0}=48\phi^{2}\chi x_{4}/\pi^{3/2}$, and\[
\eta/\eta_{0}\mbox{ or }\kappa/\kappa_{0}=\frac{c_{1}}{\sqrt{\pi}\chi x_{2}}(1+c_{2}\phi\chi x_{3})^{2}+c_{3}\eta_{B}/\eta_{0},\]
where $\zeta_{0}=D\sqrt{kT/m}$, $\eta_{0}=D^{-2}\sqrt{mkT}$ and
$\kappa_{0}=kD^{-2}\sqrt{kT/m}$ are natural units, and $c_{1}=5/48$,
$c_{2}=24/5$ and $c_{3}=3/5$ for $\eta$, while $c_{1}=25/64$,
$c_{2}=24/5$ and $c_{3}=3/5$ for $\kappa$.

\begin{figure}[tbph]
\includegraphics[width=0.85\columnwidth,keepaspectratio]{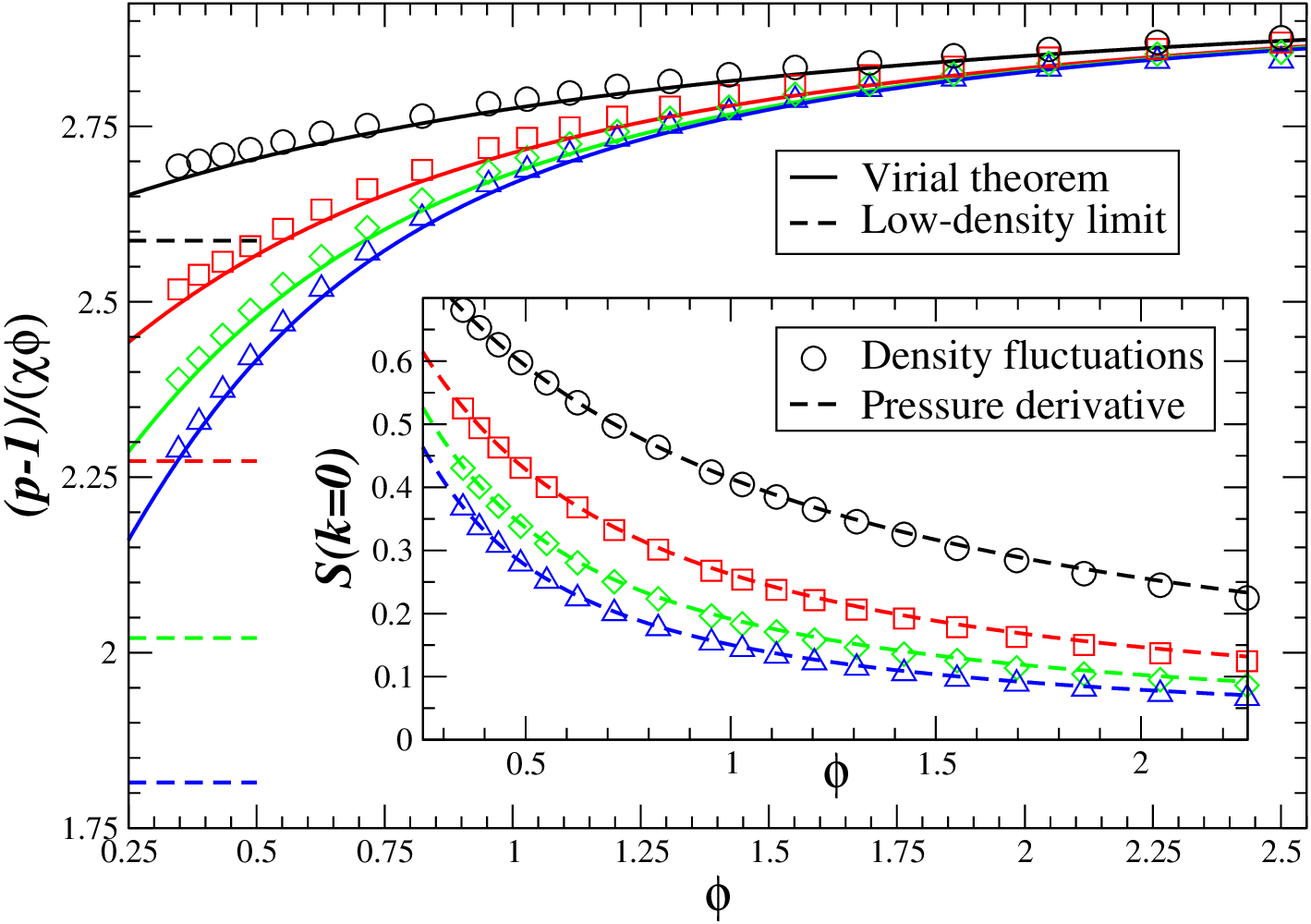}

\includegraphics[width=0.85\columnwidth,keepaspectratio]{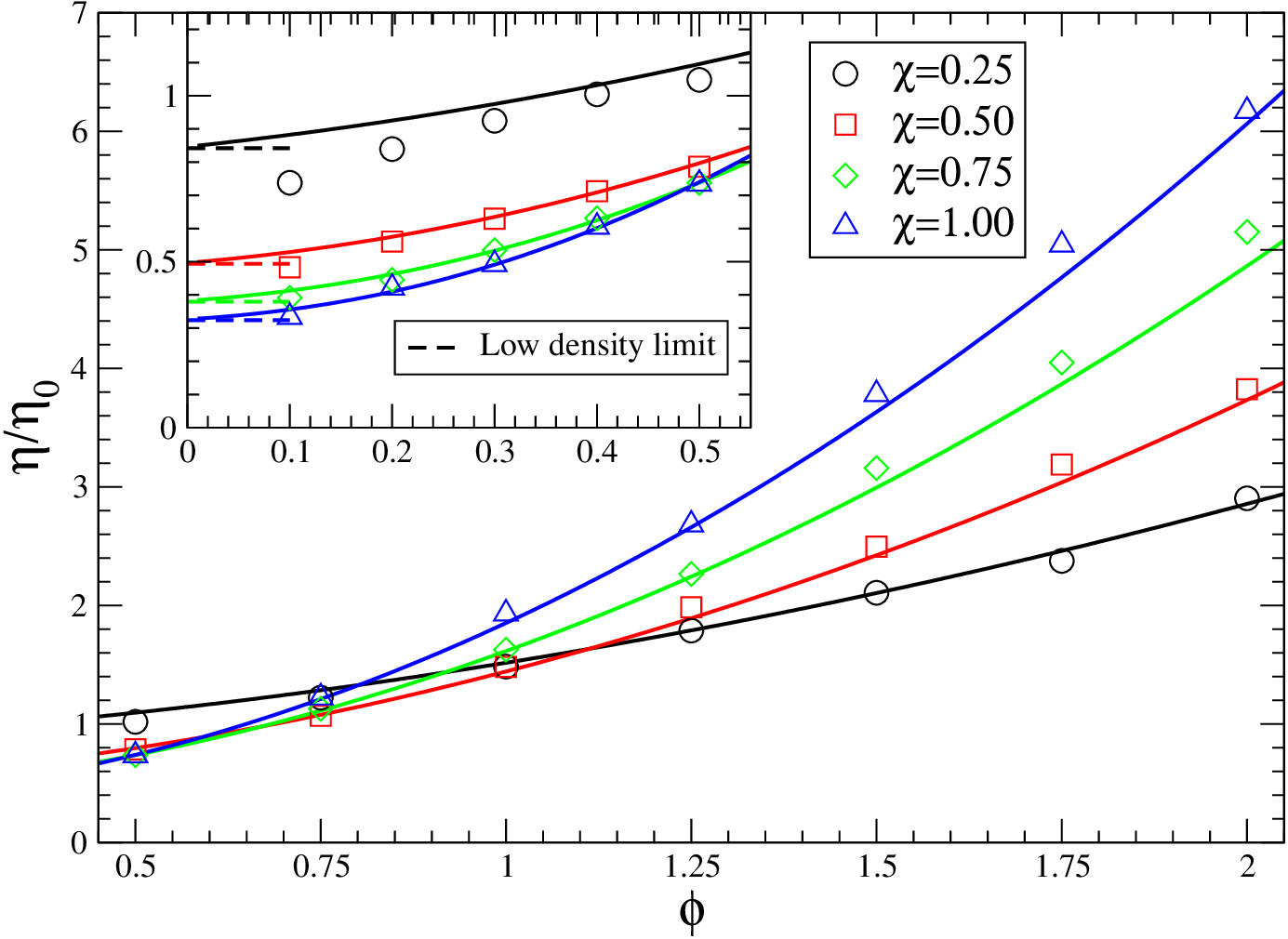}

\caption{\label{SHSD_transport_coeffs}(Color online) Comparison between numerical
results for SHSD at several collision frequencies (different symbols)
with predictions based on the stochastic Enskog equation using the
HNC $g_{2}(x)$ (solid lines). The low-density approximations are
also indicated (dashed lines). (\emph{Top}) Normalized equation of
state. The inset compares the compressibility (pressure derivative,
dashed lines) to the structure factor at the origin $S(k\rightarrow0)$
(symbols), measured using a direct Fourier transform of the particle
positions for small $k$ and extrapolating to $k=0$. (\emph{Bottom})
The shear viscosity at high and low densities (inset), as measured
using an externally-forced Poiseuille flow. There are significant
corrections (Knudsen regime) for large mean free paths (i.e., at low
densities and low collision rates).}
\end{figure}

The above formula for the pressure is exact and is equivalent to the
virial theorem for the linear core potential, and thus thermodynamic
consistency between $g_{2}(x)$ and $p(\phi)$ is guaranteed. In the
inset in the top part of Fig. \ref{SHSD_transport_coeffs}, we directly
demonstrate the thermodynamic consistency of SHSD by comparing the
compressibility calculated from the EOS, $S_{c}=(p+\phi dp/d\phi)^{-1}$,
to the structure factor at the origin $S_{0}=S(\omega=0,k=0)$. Furthermore,
good agreement is found between the adiabatic speed of sound $c_{s}^{2}=S_{0}^{-1}+2p^{2}/3$
and the location of the Brilloin lines in the dynamic structure factor
$S(\omega;k)$ for small $k$ values. In Fig. \ref{SHSD_transport_coeffs},
we also compare the theoretical predictions for $\eta$ utilizing
the HNC approximation for $g_{2}$ to the ones directly calculated
from SHSD. Surprisingly, good agreement is found for the shear viscosity
at all densities. The corresponding results for $\zeta$ show significant
($\sim25\%$) deviations for the self-diffusion coefficient at higher
densities because of corrections due to higher-order correlations.

As an illustration of the correct hydrodynamic behavior of the SHSD
fluid and the significance of compressibility, we study the velocity
autocorrelation function (VACF) $C(t)=\left\langle v_{x}(0)v_{x}(t)\right\rangle $
for a single neutrally-buoyant hard sphere of mass $m$ and radius
$R$ suspended in an SHSD fluid of mass density $\rho$. This problem
is relevant to the modeling of polymer chains or (nano)colloids in
solution, and led to the discovery of a long power-law tail in $C(t)$
\citet{BrownianSRD_Review,BrownianLB_VACF}. Here the solvent-solvent
particles interact as in SHSD. \modified{The solvent-solute interaction
is treated as if the SHSD particles are hard spheres of diameter $D_{s}$,
chosen to be somewhat smaller than their interaction diameter with
other solvent particles (specifically, we use $D_{s}=D/4$) for computational
efficiency reasons, using an event-driven algorithm \citet{DSMC_AED}.}
Upon collision the relative velocity of the solvent particle is reversed
in order to provide a no-slip condition at the surface of the suspended
sphere \citet{BrownianSRD_Review,DSMC_AED} (slip boundaries give
qualitatively identical results). For comparison, an ideal solvent
of comparable viscosity is also simulated.

Theoretically, $C(t)$ has been calculated from the linearized (compressible)
fluctuating Navier-Stokes (NS) equations \citet{BrownianSRD_Review}.
The results are analytically complex even in the Laplace domain, however,
at short times an inviscid compressible approximation applies. At
large times the compressibility does not play a role and the incompressible
NS equations can be used to predict the long-time tail. At short times,
$t<t_{c}=2R/c_{s}$, the major effect of compressibility is that sound
waves generated by the motion of the suspended particle carry away
a fraction of the momentum, so that the VACF quickly decays from its
initial value $C(0)=kT/m$ to $C(t_{c})\approx kT/M$, where $M=m+2\pi R^{3}\rho/3$.
At long times, $t>t_{visc}=4\rho R_{H}^{2}/3\eta$, the VACF decays
as in an incompressible fluid, with an asymptotic power-law tail $(kT/m)(8\sqrt{3\pi})^{-1}(t/t_{visc})^{-3/2}$,
in disagreement with predictions based on the Langevin equation (Brownian
dynamics), $C(t)=(kT/m)\exp\left(-6\pi R_{H}\eta t/m\right)$. We
have estimated the effective (hydrodynamic) colloid radius $R_{H}$
from numerical measurements of the Stokes friction force $F=-6\pi R_{H}\eta v$.

\begin{figure}[tbph]
\includegraphics[width=1\columnwidth,keepaspectratio]{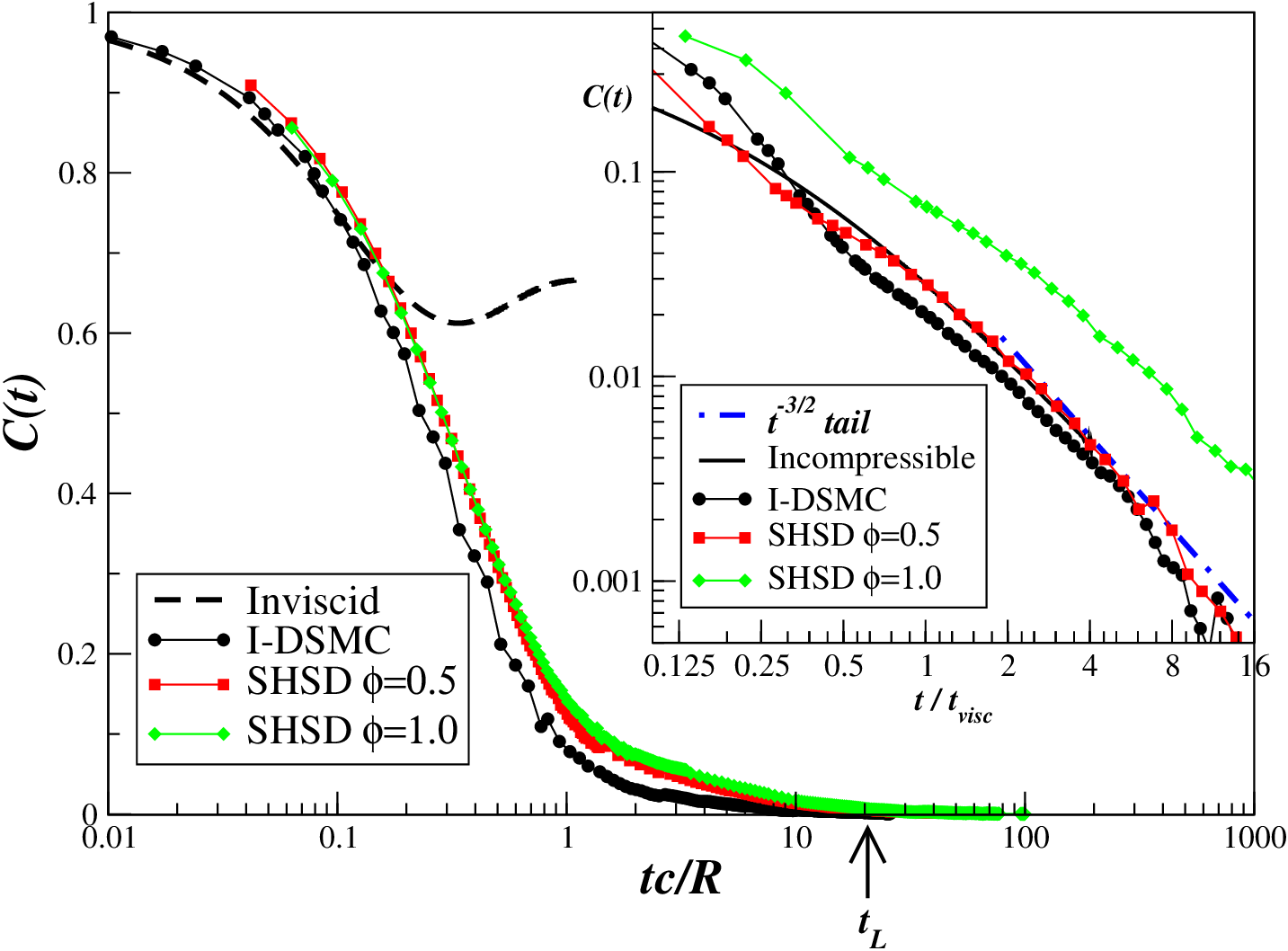}

\caption{\label{VACF_SHSD}(Color online) The velocity autocorrelation function
for a neutrally buoyant hard sphere suspended in a non-ideal SHSD
($\chi=1$) solvent at two densities (symbols), as well as an ideal
I-DSMC solvent ($\phi=0.5$, $\chi=0.62$, symbols), at short and
long times (inset). \modified{For the more compressible (less viscous)
fluids the long time tails are statistically measurable only up to
$t/t_{visc}\approx5$.} The theoretical predictions based on the
inviscid, for short times, or incompressible, for long times, Navier-Stokes
equations are also shown (lines). The diameter of the nano-colloidal
particle is only $2.5D$, although we have performed simulations using
larger spheres as well with very similar results. Since periodic boundary
conditions were used we only show the tail up to about the time at
which sound waves generated by its periodic images reach the particle,
$t_{L}=L/c_{s}$.}
\end{figure}

In Fig. \ref{VACF_SHSD} numerical results for the VACF for an I-DSMC
solvent and an SHSD solvent at two different densities are compared
to the theoretical predictions. It is seen, as predicted, that the
compressibility or the sound speed $c_{s}$, determines the early
decay of the VACF. The exponent of the power-law decay at large times
is also in agreement with the hydrodynamic predictions. The coefficient
of the VACF tail agrees reasonably well with the hydrodynamic prediction
for the less dense solvents, however, there is a significant deviation
of the coefficient for the densest solvent, perhaps due to ordering
of the fluid around the suspended sphere, not accounted for in continuum
theory.

\deleted{We have successfully designed a thermodynamically-consistent
DSMC-like algorithm for non-ideal fluids. }\added{In closing, we
should point out that for reasonable values of the collision frequency
($\chi\sim1$) and density ($\phi\sim1$) the SHSD fluid is still
relatively compressible compared to a dense liquid, $c_{s}^{2}<10$.
Indicative of this is that the diffusion coefficient is large relative
to the viscosity as in typical DPD simulations, so that the Schmidt
number $S_{c}=\eta(\rho\zeta)^{-1}$ is less than 10 instead of being
on the order of 100-1000. Achieving higher $c_{s}$ or $S_{c}$ requires
high collision rates (for example, $\chi\sim10^{4}$ is used in Ref.
\citet{DPDSchmidtNumbers}) and appropriately smaller time steps to
ensure that there is at most one collision per particle per time step,
and thus a similar computational effort as in molecular dynamics.
The advantage of SHSD is its simplicity, easy parallelization, and
simpler coupling to continuum methods such as fluctuating hydrodynamics
\citet{FluctuatingHydro_AMAR}.}

This work performed under the auspices of the U.S. Department of Energy
by Lawrence Livermore National Laboratory under Contract DE-AC52-07NA27344
(LLNL-JRNL-401745). We thank Salvatore Torquato, Frank Stillinger,
Ard Louis, Andres Santos, and Jacek Polewczak for their assistance
and advice.


\end{document}